\documentclass[fleqn]{article}
\usepackage[centertags]{amsmath}
\usepackage{espcrc2}
\usepackage{graphicx}
\title{External Field Dependence of Deconfinement
          Temperature in SU(3)}

\author{Paolo Cea\address[DF,INFN]{Dipartimento di Fisica, Univ. of Bari and INFN - Sezione di Bari,
        I-70126 Bari, Italy}  and
        Leonardo Cosmai\address[INFN]{INFN - Sezione di Bari, I-70126 Bari, Italy}}

\begin{document}

\begin{abstract}
We study vacuum dynamics of SU(3) lattice gauge theory at finite
temperature.  Using the lattice Schr\"odinger functional, SU(3) vacuum
is probed by means of an external constant Abelian chromomagnetic
field.  Our preliminary numerical data suggest that, by increasing the
strength of the applied external field, deconfinement temperature
decreases towards zero.  This means that strong enough Abelian
chromomagnetic fields destroy confinement of color.
\vspace{1pc}
\end{abstract}

\maketitle

To study vacuum structure of lattice gauge theories we
introduced~\cite{Cea:1997ff,Cea:1999gn} a gauge invariant
effective action, defined by means of the lattice Schr\"odinger
functional
\begin{equation}
\label{Zetalatt}
{\mathcal{Z}}[U^{\mathrm{ext}}_\mu] = \int {\mathcal{D}}U \; e^{-S_W} \,.
\end{equation}
$S_W$ is the standard
Wilson action and the functional
integration is extended over links on a lattice
$L_s^3 \times L_t$  with the
hypertorus geometry  and satisfying the
constraints ($x_t$: temporal coordinate)
\begin{equation}
\label{coldwall} U_k(x)|_{x_t=0} = U^{\mathrm{ext}}_k(x)
\,,\,\,\,\,\, (k=1,2,3) \,\,,
\end{equation}
$U^{\mathrm{ext}}_k(x)$ being the lattice version of
the external continuum gauge field
$\vec{A}^{\mathrm{ext}}(x)=
\vec{A}^{\mathrm{ext}}_a(x) \lambda_a/2$.

The lattice effective action
for the external static background field
$\vec{A}^{\mathrm{ext}}(x)$ is given by
\begin{equation}
\label{Gamma} \Gamma[\vec{A}^{\mathrm{ext}}] = -\frac{1}{L_4} \ln
\left\{
\frac{{\mathcal{Z}}[\vec{A}^{\mathrm{ext}}]}{{\mathcal{Z}}[0]}
\right\}
\end{equation}
and is invariant for  gauge transformations
of the external links $U^{\mathrm{ext}}_k$.

At finite temperature $T=1/a L_t$, we introduced the thermal partition
function in presence of a given static background field:
\begin{equation}
\begin{split}
\label{ZetaTnew} & \mathcal{Z}_T \left[ \vec{A}^{\text{ext}}
\right]
= \\
& \qquad \qquad
\int_{U_k(L_t,\vec{x})=U_k(0,\vec{x})=U^{\text{ext}}_k(\vec{x})}
\mathcal{D}U \, e^{-S_W}   \,.
\end{split}
\end{equation}
In this case the relevant quantity is the free energy functional
\begin{equation}
\label{freeenergy} F[\vec{A}^{\mathrm{ext}}]= - \frac{1}{L_t} \ln
\frac{\mathcal{Z}_T[\vec{A}^{\mathrm{ext}}]} {{\mathcal{Z}}_T[0]}
\,.
\end{equation}
If physical temperature is sent to zero the thermal functional
Eq.~(\ref{ZetaTnew}) reduces to Eq.~(\ref{Zetalatt}).
We used the above defined lattice effective action to investigate
if deconfinement temperature depends on the strength of an applied
external constant Abelian chromomagnetic field.

The static constant Abelian chromomagnetic field in
the continuum is
\begin{equation}
\label{field}
\vec{A}^{\mathrm{ext}}_a(\vec{x}) =
\vec{A}^{\mathrm{ext}}(\vec{x}) \delta_{a,3} \,, \quad
A^{\mathrm{ext}}_k(\vec{x}) =  \delta_{k,2} x_1 H \,.
\end{equation}
On the lattice
\begin{equation}
\label{t3links}
\begin{split}
& U^{\mathrm{ext}}_1(\vec{x}) =
U^{\mathrm{ext}}_3(\vec{x}) = {\mathbf{1}} \,,
\\
& U^{\mathrm{ext}}_2(\vec{x}) =
\begin{bmatrix}
\exp(i \frac {g H x_1} {2})  & 0 & 0 \\ 0 &  \exp(- i \frac {g H
x_1} {2}) & 0
\\ 0 & 0 & 1
\end{bmatrix}
\end{split}
\end{equation}
Due to periodic boundary conditions the magnetic field $H$ turns
out to be quantized
\begin{equation}
\label{quant} \frac{a^2 g H}{2} = \frac{2 \pi}{L_1}
n_{\mathrm{ext}} \,, \qquad  n_{\mathrm{ext}}\,\,\,{\text{integer}}\,.
\end{equation}
This field gives rise to constant field strength, then due to
gauge invariance it is easy to show that
$F[\vec{A}^{\mathrm{ext}}]$ is proportional to spatial volume
$V=L_s^3$. Therefore the relevant quantity is the density
$f[\vec{A}^{\mathrm{ext}}]$ of  free energy
\begin{equation}
\label{free-energy} f[\vec{A}^{\mathrm{ext}}] = \frac{1}{V}
F[\vec{A}^{\mathrm{ext}}] \,.
\end{equation}
The $\beta$-derivative of $f[\vec{A}^{\mathrm{ext}}]$ (at fixed
external field strength $gH$)  is easily evaluated ($\Omega =
L_s^3 \times L_t$)
\begin{equation}
\label{deriv}
\begin{split}
f^{\prime}[\vec{A}^{\mathrm{ext}}]   & = \left \langle
\frac{1}{\Omega} \sum_{x \in \tilde{\Lambda},\mu < \nu}
\frac{1}{3} \text{Re} {\text{Tr}} U_{\mu\nu}(x) \right\rangle_0  \\
& - \left\langle \frac{1}{\Omega} \sum_{x \in \tilde{\Lambda},\mu
< \nu} \frac{1}{3} \text{Re} {\text{Tr}} U_{\mu\nu}(x)
\right\rangle_{\vec{A}^{\mathrm{ext}}} \,.
\end{split}
\end{equation}
The generic plaquette
$U_{\mu\nu}(x)=U_\mu(x)U_\nu(x+\hat{\mu})U^\dagger_\mu(x+\hat{\nu})U^\dagger_\nu(x)$
contributes to the sum in Eq.~(\ref{deriv}) if the link $U_\mu(x)$
is a "dynamical" one (i.e. it is not constrained in the functional
integration Eq.~(\ref{ZetaTnew})). $\tilde{\Lambda}$ is a
sub-ensemble of the lattice sites, $x \in \tilde{\Lambda}$ if the
link $U_\mu(x)$ exiting from it is a dynamical link.

As is well known pure SU(3) gauge system undergoes a deconfinement
phase transition by increasing temperature. To estimate
deconfinement temperature $T_c$, we evaluate
$f^{\prime}[\vec{A}^{\mathrm{ext}}]$ as a function of $\beta$ for
different lattice temporal sizes $L_t$. To determine the
pseudocritical couplings we parameterize $f^{\prime}(\beta,L_t)$
(see Fig.~1) near the peak as
\begin{equation}
\label{peak-form}
\frac{f^{\prime}(\beta,L_t)}{\varepsilon^{\prime}_{\mathrm{ext}}}
= \frac{a_1(L_t)}{a_2(L_t) [\beta - \beta^*(L_t)]^2 +1} \,.
\end{equation}
In Eq.~(\ref{peak-form}) $\varepsilon^{\prime}_{\mathrm{ext}}$ is
the  derivative of the classical energy due to  the external
applied field
\begin{equation}
\label{epsprimeext} \varepsilon^{\prime}_{\mathrm{ext}} =
\frac{2}{3} \, [1 - \cos( \frac{g H}{2} )] = \frac{2}{3} \, [1 -
\cos( \frac{2 \pi}{L_1} n_{\mathrm{ext}})] .
\end{equation}
\begin{figure}[!ht]
\begin{center}
\includegraphics[width=0.4\textwidth,clip]{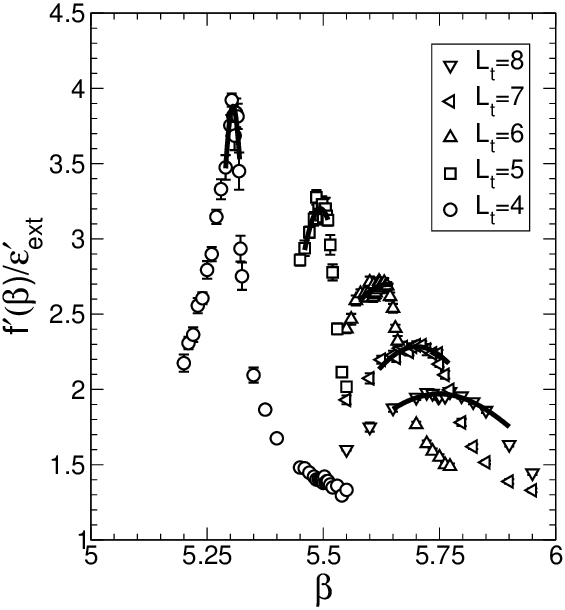}
\vspace{-1.1cm} \caption{$f^{\prime}$ versus $\beta$ for various
$L_t$ at given external field strength ($n_{\text{ext}}=1$.)}
\end{center}
\end{figure}
Once determined $\beta^*(L_t)$ we estimate the deconfinement
temperature as
\begin{equation}
\label{Tc} \frac{T_c}{\Lambda_{\mathrm{latt}}} = \frac{1}{L_t}
\frac{1}{f(\beta^*(L_t))} \,,
\end{equation}
with
\begin{equation}
\label{af} f(\beta) = \left( \frac{\beta}{2  N b_0}
\right)^{b_1/2b_0^2} \, \exp \left( -\beta \frac{1}{4 N b_0} \right)
\,,
\end{equation}
where $N$ is the color number, $b_0=(11 N)/(48 \pi^2)$, $b_1=(34N^2)/(3(16\pi^2)^2)$.

Following~\cite{Fingberg:1993ju} we perform a linear extrapolation
to the continuum of our data for $T_c/\Lambda_{\mathrm{latt}}$. We
see that our estimate of $T_c/\Lambda_{\mathrm{latt}}$ in the
continuum is in fair agreement with the one available in the
literature~\cite{Fingberg:1993ju} without external field (see
Fig.~2).
\begin{figure}[!ht]
\begin{center}
\includegraphics[width=0.4\textwidth,clip]{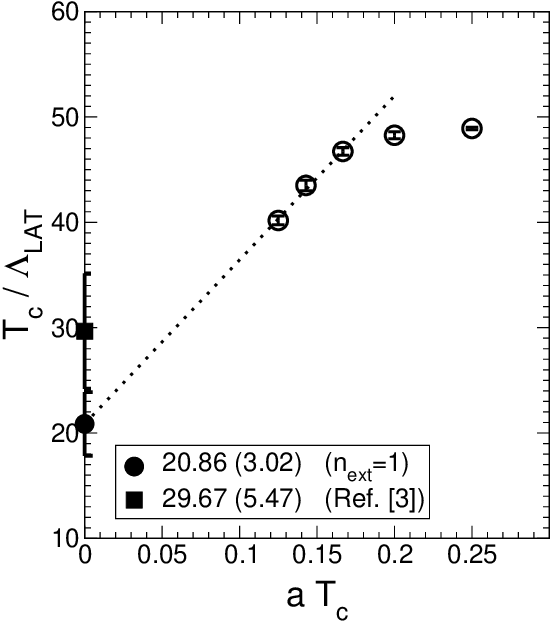}
\vspace{-1.1cm}  \caption{The estimate of $T_c$ for
$n_{\text{ext}}=1$ and $n_{\text{ext}}=0$.}
\end{center}
\end{figure}
However Fig.~2 is suggestive of a possible non-trivial dependence
of deconfinement critical temperature on  external Abelian
chromomagnetic field.  Therefore we decided to vary the strength
of the applied external Abelian chromomagnetic background field to
study possible dependence of $T_c$ on $gH$. To this aim we
performed numerical simulations on $64^3 \times L_t$ lattices with
$n_{\text{ext}}=1,5,10$.
In Fig.~3 we display our determination of $T_c$ for two different
values ($n_{\text{ext}}=1,10$) of the applied field strength. We
see that the critical temperature decreases by increasing the
external Abelian chromomagnetic field. For dimensional reasons one
expects that
\begin{equation}
\label{magcrit} T_c^2 \, \, \sim \,  \, gH \, .
\end{equation}
\begin{figure}[!ht]
\begin{center}
\includegraphics[width=0.4\textwidth,clip]{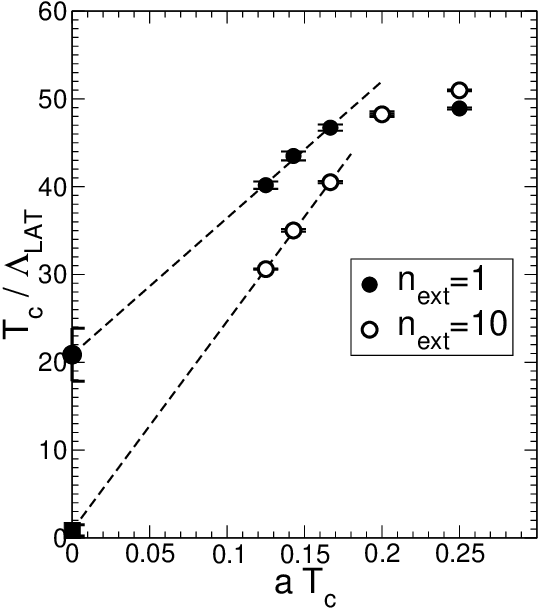}
\vspace{-1.1cm} \caption{The estimate of $T_c$ for
$n_{\text{ext}}=1$ and $n_{\text{ext}}=10$.}
\end{center}
\end{figure}
\begin{figure}[!ht]
\begin{center}
\includegraphics[width=0.4\textwidth,clip]{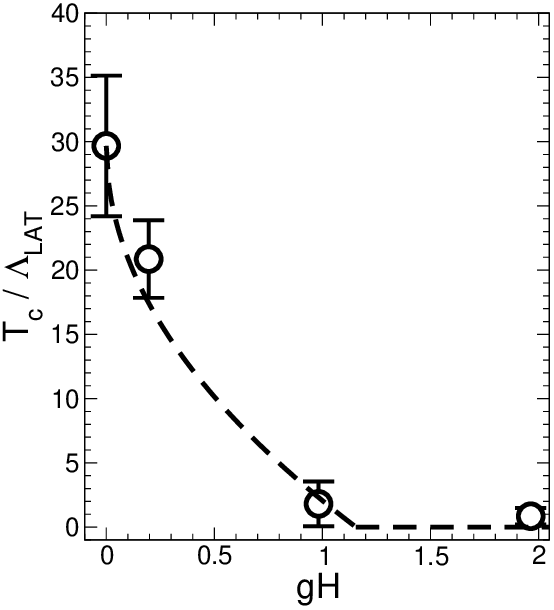}
\vspace{-0.95cm} \caption{$T_c/\Lambda_{\text{LAT}}$ versus the
applied external field strength $gH$.}
\end{center}
\end{figure}
Indeed we try to fit our data with
\begin{equation}
\label{Tcfit} \frac{T_c(gH)}{\Lambda_{\text{latt}}} =
\frac{T_c(0)}{\Lambda_{\text{latt}}} + \alpha \sqrt{gH} \,.
\end{equation}
We find a satisfying fit with $\alpha=-27.6 \pm 2.6$ (see Fig.~4).
Remarkably, we see that there exists a critical field
\begin{equation}
\label{Hc} a^2 gH_c \simeq 1.15
\end{equation}
such that $T_c=0$ for $H>H_c$.

In conclusion, we found~\cite{Cea:1999gn} that at
zero-temperature, in the thermodynamic limit, SU(3) vacuum screens
completely external chromomagnetic Abelian field. Moreover, by
increasing temperature there is no screening of the Abelian
chromomagnetic field. There is a critical field $H_c$ such that,
for $H>H_c$, the gauge system is in the deconfined phase. In
particular, this last result suggests that there is an intimate
connection between Abelian chromomagnetic fields and color
confinement.


\begin{thebibliography}{1}

\bibitem{Cea:1997ff}
P. Cea, L. Cosmai and A.D. Polosa,
\newblock Phys. Lett. B392 (1997) 177, hep-lat/9601010.

\bibitem{Cea:1999gn}
P. Cea and L. Cosmai,
\newblock Phys. Rev. D60 (1999) 094506, hep-lat/9903005.

\bibitem{Fingberg:1993ju}
J. Fingberg, U. Heller and F. Karsch,
\newblock Nucl. Phys. B392 (1993) 493, hep-lat/9208012.

\end{thebibliography}

\end{document}